\documentclass[journal]{IEEEtran}

\usepackage{amsmath,amssymb,amsfonts,color}
\usepackage{graphicx}
\usepackage{caption}
\usepackage{algorithm}
\usepackage{algorithmic}
\usepackage{cite}
\usepackage{bm}
\usepackage[none]{hyphenat}
\usepackage{hyperref}
\newcommand{\C}{\mathbb{C}}

\newcommand{\diag}{\operatorname{diag}}
\newcommand{\blkdiag}{\operatorname{blkdiag}}
\newcommand{\CN}{\mathcal{CN}}
\newcommand{\kr}{\diamond}
\newcommand{\ten}[1]{\boldsymbol{\mathcal{#1}}}
\newtheorem{remark}{Remark}

\title{TRACE: A Trilinear Channel Estimation Framework for Tri-Hybrid Architectures with Reconfigurable Antennas}

\author{André L. F. de Almeida,~\IEEEmembership{Senior Member,~IEEE,}
        and~Mengzhen Liu
\thanks{André L. F. de Almeida is with the Laboratory of Applied Signal Processing and the Wireless Telecom Research Group, Department of Teleinformatics Engineering, Federal University of Ceará, Fortaleza, Brazil. E-mail: andre@gtel.ufc.br. This work is partially supported by the National Institute of Science and Technology (INCT-Signals) sponsored by Brazil's National Council for Scientific and Technological Development (CNPq) (406517/2022-3), and FUNCAP (INCT-25255-82587.32.41/64). The research of André L. F. de Almeida is partially supported by CNPq (303356/2025-1). 
}
\thanks{Mengzhen Liu is with the School of Information and Communication Engineering, Dalian University of Technology, Dalian 116024, China (e-mail: liumengzhen@mail.dlut.edu.cn).}}

\renewcommand\baselinestretch{.99}

\begin{document}
\raggedbottom
\maketitle

\begin{abstract}
Tri-hybrid beamforming with reconfigurable antennas (RAs) adds an electromagnetic (EM)-domain degree of freedom to radio-frequency (RF) analog combining and baseband processing, but it also couples the EM angular response, the spatial steering matrix, and the path gains inside every received pilot. This letter develops TRACE, a trilinear channel estimation framework for the single-user uplink of a tri-hybrid receiver. Cycling the radiation pattern over the pilot dimension yields a third-order PARAFAC tensor of the received pilots, with factors corresponding to the EM-domain angular factor, the RF-projected steering factor, and the pilot-weighted fading factor. TRACE fits this tensor by alternating least squares and then recovers the physical channel parameters in closed form. Kruskal-based identifiability translates into design rules for the number of pilot slots, RF chains, and EM probing states. In a compressive configuration, TRACE attains the SNR slope of an oracle-support bound, staying within a factor of $1.4$ of it above $10$~dB, while outperforming a dictionary-based greedy baseline.
\end{abstract}

\begin{IEEEkeywords}
Channel estimation, reconfigurable antennas, tri-hybrid beamforming, PARAFAC decomposition.
\end{IEEEkeywords}

\renewcommand{\baselinestretch}{.88}

\section{Introduction}

Pattern reconfigurable antennas (RAs) electronically reshape their radiation pattern, adding an electromagnetic (EM)-domain degree of freedom (DoF) beyond that of fixed-pattern arrays~\cite{murchPatternAntenna,boerman2008pattern}. Pixel- and parasitic-based designs realize large discrete pattern sets in practice~\cite{towfiq2018reconfigurable}, and combining them with hybrid beamforming yields a \emph{tri-hybrid} architecture, in which radiation patterns are selected in the EM domain, phase-coherent combining is performed in the radio-frequency (RF) domain, and digital processing is carried out in the baseband (BB) domain~\cite{heath2026trihybrid,castellanos2026embracing}.
Tri-hybrid beamforming has been shown to improve spectral efficiency, energy efficiency, and hardware scalability across a range of applications~\cite{liu2026tritimescale,chenTriHybridISAC,zheng2025trihybridprecoding}, and the associated signal-processing primitives have recently been formalized~\cite{deshpande2026spfoundations,raArraysOverview}. Realizing these gains presupposes accurate acquisition of the multi-domain channel that the EM, RF, and BB beamformers act upon.

Channel estimation for hybrid arrays is classically posed as a sparse recovery problem over an angular dictionary and solved by greedy simultaneous orthogonal matching pursuit (SOMP)~\cite{alkhateeb2014channel}. For reconfigurable architectures, training design and estimation with mode-switching RAs were studied in~\cite{bahceci2017efficient}, EM-domain estimation for reconfigurable massive MIMO in~\cite{ying2025reconfigurable}, and a tri-domain scheme combining pattern reconfiguration with antenna repositioning in~\cite{li2026tridomain}. These estimators are either dictionary-based or parametric, and they treat the EM, spatial, and fading components sequentially. However, they do not exploit the multilinear coupling induced by pattern reconfiguration. In different settings, tensor decompositions have proven to be a natural way to estimate the channels of reconfigurable surfaces~\cite{araujo2021tensor,wei2021parafac}, suggesting that the same structure should be exploited here.

Accordingly, this letter proposes TRACE, and to the best of our knowledge, it is the first tensor-based channel estimator for tri-hybrid architectures with pattern RAs. The contributions are as follows. First, a trilinear signal model is derived that preserves the coupling among the EM radiation pattern, the RF-projected steering response, and the BB path gains, representing the received pilots as a third-order PARAFAC tensor. Second, a two-stage estimator is developed that fits the tensor by alternating least squares (ALS) and then recovers the physical parameters. Third, Kruskal's uniqueness condition is translated into explicit design rules on the pilot, RF, and EM probing dimensions, and is validated numerically against a greedy dictionary baseline and an oracle-support bound.

\section{System and Tensor Model}\label{sec:model}

\begin{figure}[!t]
  \centering
  \includegraphics[width=3.3in]{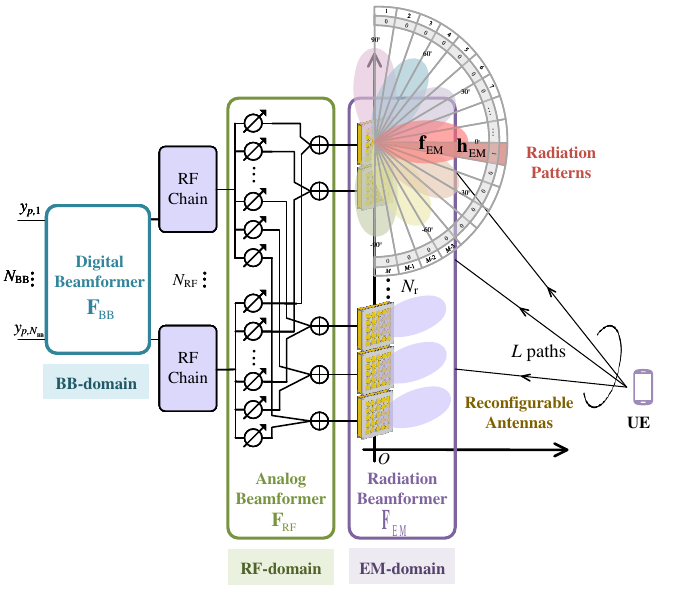}
  \vspace{-0.0cm}
  \caption{Tri-hybrid beamforming architecture.}
  \vspace{-0.4cm}
  \label{fig:systemmodel}
\end{figure}

Consider a single-user uplink SIMO system in which the base station (BS) employs the tri-hybrid architecture shown in Fig.~\ref{fig:systemmodel}, with $N_{\rm r}$ pattern RAs and $N_{\rm RF}$ RF chains serving a single-antenna user equipment (UE). \textcolor{black}{Training uses $P$ pilot slots and $T$ EM probing periods per slot, with the known symbol $s_p$ repeated over the $T$ periods of slot $p$.} Following~\cite{liu2026tritimescale}, the EM-domain radiation beamformer is
\begin{equation}
    \mathbf F_{\rm EM}=\blkdiag\{\mathbf f_{{\rm EM},1}^{T},\ldots,
    \mathbf f_{{\rm EM},N_{\rm r}}^{T}\}\in\C^{N_{\rm r}\times MN_{\rm r}},
    \label{eq:FEM}
\end{equation}
where $\mathbf{f}_{\mathrm{EM},n}\in\C^{M}$ is the complex far-field response of the $n$-th RA in its current reconfiguration state, sampled at $M$ angular samples; $\mathbf F_{\rm RF}\in\C^{N_{\rm RF}\times N_{\rm r}}$ has constant-modulus entries realized by phase shifters, and $\mathbf F_{\rm BB}$ performs digital combining. Note that $\mathbf F_{\rm RF}$ and $\mathbf F_{\rm EM}$ are objects of different kinds: the former is a commanded \emph{weight} matrix, whereas the latter is a \emph{characterization} of the array obtained offline by measurement or EM simulation. \textcolor{black}{The far-field response is generally complex-valued because it contains both amplitude and phase information; its squared magnitude gives the real, non-negative radiated gain.}
The spatial channel at slot $p$ is the superposition of $L$ propagation paths,
\begin{equation}
    \mathbf h_{p}=\sum_{\ell=1}^{L}\alpha_{p,\ell}\,\mathbf a(\theta_{\ell})
    \triangleq\mathbf A(\bm\theta)\bm\alpha_{p},
    \label{eq:hp}
\end{equation}
where $\mathbf{A}(\bm\theta)\triangleq[\mathbf a(\theta_{1}),\ldots,\mathbf a(\theta_{L})]\in\C^{N_{\mathrm{r}}\times L}$ collects the steering vectors of the angles of arrival (AoAs), assumed constant over the $P$ slots, and $\bm\alpha_{p}\triangleq[\alpha_{p,1},\ldots,\alpha_{p,L}]^{T}\in\C^{L}$ collects the fading coefficients, which vary with $p$. Pattern RAs extend this to an $M$-dimensional representation carrying the EM angular index, as follows
\begin{equation}
    \overline{\mathbf{h}}_{p}
    =\sum_{\ell=1}^{L}\alpha_{p,\ell}
    (\mathbf a(\theta_{\ell})\otimes\mathbf h_{{\rm EM},\ell})
    =(\mathbf{A}(\bm\theta)\kr\overline{\mathbf H}_{\mathrm{EM}})\bm\alpha_{p},
    \label{eq:hbar}
\end{equation}
where $\kr$ is the Khatri--Rao product and $\overline{\mathbf H}_{\mathrm{EM}}=[\mathbf h_{{\rm EM},1},\ldots,\mathbf h_{{\rm EM},L}]\in\{0,1\}^{M\times L}$ has one-hot columns indicating the EM angular sample of each path. The BB-combined received pilot at slot $p$ is given by
\begin{equation}
    \mathbf{y}_{p}
    =\mathbf F_{\rm BB}\mathbf F_{\rm RF}\mathbf F_{\rm EM}
    \left(\mathbf A(\bm\theta)\kr\overline{\mathbf H}_{\rm EM}\right)
    \bm\alpha_{p}\,s_{p}+\mathbf{n}_{p},
    \label{eq:rx1}
\end{equation}
with $s_{p}$ the known pilot symbol, \textcolor{black}{$|s_{p}|^{2}=1/(PT)$ so that the total pilot energy over all $PT$ transmissions is unity}, and $\mathbf n_{p}$ the noise after combining.

\emph{Tensor formulation.} TRACE keeps the RF-chain observations available before digital combining, i.e.\ $\mathbf F_{\rm BB}=\mathbf I_{N_{\rm RF}}$ during training. All $N_{\rm r}$ antennas share the same pattern in each EM probing state, so at probing period $t$ the array-level radiation beamformer is
\begin{equation}
    \mathbf F_{{\rm EM},t}=\mathbf I_{N_{\mathrm{r}}}\otimes
    \overline{\mathbf f}_{{\rm EM},t}^{T},
    \label{eq:common_em_pattern}
\end{equation}
where the BS cycles through $T$ reconfiguration states $\{\overline{\mathbf f}_{{\rm EM},1},\ldots,\overline{\mathbf f}_{{\rm EM},T}\}$ within each slot and repeats the same cycle across slots. This common-pattern assumption is what lets the EM training act on the angular mode $\overline{\mathbf H}_{\rm EM}$ in a structured way. Inserting \eqref{eq:common_em_pattern} into \eqref{eq:rx1} and applying the mixed-product rule, the RF/EM-projected channel collapses to a diagonal,
\begin{equation}
    \mathbf F_{\rm RF}\mathbf F_{{\rm EM},t}
    \left(\mathbf A(\bm\theta)\kr\overline{\mathbf H}_{\rm EM}\right)
    =\mathbf F_{\rm RF}\mathbf A(\bm\theta)
    \diag\big(\overline{\mathbf f}_{{\rm EM},t}^{T}
    \overline{\mathbf H}_{\rm EM}\big).\nonumber
\end{equation}
This step creates the trilinear structure: the EM state index $t$ enters only through a diagonal scaling of the $L$ paths. Hence
\begin{equation}
    \mathbf y_{t,p}=
    \mathbf F_{\rm RF}\mathbf A(\bm\theta)
    \diag\big(\overline{\mathbf f}_{{\rm EM},t}^{T}
    \overline{\mathbf H}_{\rm EM}\big)
    \bm\alpha_{p}\,s_p+
    \mathbf n_{t,p}.
    \label{eq:mode_absorption}
\end{equation}
Collecting the received pilots over the $T$ probing states and the $P$ slots into the matrix
\begin{equation}
    \mathbf Y=
    \begin{bmatrix}
    \mathbf{y}_{1,1} & \cdots & \mathbf{y}_{1,P}\\
    \vdots & \ddots & \vdots\\
    \mathbf{y}_{T,1} & \cdots & \mathbf{y}_{T,P}
    \end{bmatrix}\in\C^{TN_{\rm RF}\times P},
    \label{eq:Ymat}
\end{equation}
and using the identity $[\mathbf A\diag(\mathbf B_{1,:});\ldots;\mathbf A\diag(\mathbf B_{J,:})]=\mathbf B\kr\mathbf A$ with $(\mathbf A,\mathbf B)\leftrightarrow(\mathbf F_{\rm RF}\mathbf A(\bm\theta),\overline{\mathbf F}_{\rm EM}\overline{\mathbf H}_{\rm EM})$, yields
\begin{equation}
    \mathbf Y=
    \left[
    \left(\overline{\mathbf F}_{\rm EM}\overline{\mathbf H}_{\rm EM}\right)
    \kr\left(\mathbf F_{\rm RF}\mathbf A(\bm\theta)\right)
      \right]\mathbf B^T+\mathbf N,
    \label{eq:matrix_form}
\end{equation}
where the EM probing matrix and the pilot-weighted gain matrix are given by
\begin{align}
    \overline{\mathbf F}_{\rm EM}
    &=\left[\overline{\mathbf f}_{{\rm EM},1},\ldots,
    \overline{\mathbf f}_{{\rm EM},T}\right]^T\in\C^{T\times M},\label{eq:Fem}\\
    \mathbf B
    &=\left[\bm\alpha_{1}s_1,\ldots,
    \bm\alpha_{P}s_P\right]^T=\diag(\mathbf s)\,\mathbf C\in\C^{P\times L},\label{eq:Bdef}
\end{align}
with $\mathbf s=[s_1,\ldots,s_P]^{T}$ and $\mathbf C=[\bm\alpha_{1},\ldots,\bm\alpha_{P}]^{T}$ stacking the fading coefficients over all slots. The signal part of \eqref{eq:matrix_form} is the transpose of the mode-3 unfolding of a third-order PARAFAC model of the pilot tensor $\ten{Y}\in\C^{N_{\rm RF}\times T\times P}$,
\begin{equation}
    \ten{Y}
    =\ten{I}_{3,L}\times_1\mathbf T_A
    \times_2\mathbf T_{H}\times_3\mathbf T_{S}
    +\ten{N},
    \label{eq:parafac}
\end{equation}
whose factors are physically meaningful,
\begin{align}
    \mathbf T_{A}&=\mathbf F_{\rm RF}\mathbf A(\bm\theta)\in\C^{N_{\rm RF}\times L},\label{eq:TAeq}\\
    \mathbf T_H&=\overline{\mathbf F}_{\rm EM}\overline{\mathbf H}_{\rm EM}\in\C^{T\times L},\label{eq:THeq}\\
    \mathbf T_{S}&=\mathbf B=\diag(\mathbf s)\,\mathbf C\in\C^{P\times L},\label{eq:TSeq}
\end{align}
referred to as the RF-projected spatial steering factor, the EM-domain angular factor, and the pilot-weighted fading factor, respectively. Equivalently, in outer-product notation,
\begin{equation}
    \ten{Y}=\sum_{\ell=1}^{L}
    \mathbf t_{A,\ell}\circ\mathbf t_{H,\ell}\circ\mathbf t_{S,\ell}
    +\ten{N},
    \label{eq:rank1sum}
\end{equation}
where $\circ$ denotes the outer product and $\mathbf t_{X,\ell}$ is the $\ell$-th column of $\mathbf T_{X}$: each propagation path contributes exactly one rank-one term, so the tensor rank equals the number of resolvable paths. Since the pilot is a diagonal scaling, recovering $\mathbf C$ from $\mathbf B$ is simply division by a known scalar.

Since $\overline{\mathbf F}_{\rm EM}$ characterizes the array, designing it means selecting which reconfiguration states to cycle through. The unit-modulus independent uniform-phase model used for $\overline{\mathbf f}_{{\rm EM},t}$ in Section~\ref{sec:results} is accordingly a \emph{statistical surrogate} for a sufficiently diverse set of such states, adopted because it yields low mutual coherence among the columns of $\overline{\mathbf F}_{\rm EM}$,
\begin{equation}
    \mu\big(\overline{\mathbf F}_{\rm EM}\big)=
    \max_{m\neq m'}
    \frac{\big|\overline{\mathbf f}_{{\rm EM},m}^{H}\overline{\mathbf f}_{{\rm EM},m'}\big|}
         {\big\|\overline{\mathbf f}_{{\rm EM},m}\big\|_2
          \big\|\overline{\mathbf f}_{{\rm EM},m'}\big\|_2},
    \label{eq:coherence}
\end{equation}
on which the support detection of Section~\ref{sec:postproc} relies. It is not a claim of physical realizability: an aperture's angular response is the Fourier transform of its current distribution, so neighbouring samples are correlated. 

\section{The TRACE Estimator}\label{sec:trace}

\subsection{Tensor factor estimation}
The factors are estimated by solving the PARAFAC fitting problem
\vspace{-2ex}
\begin{equation}
    \min_{\mathbf T_A,\mathbf T_{H},\mathbf T_{S}}
    \Big\|\ten{Y}-\ten{I}_{3,L}\times_1\mathbf T_A
    \times_2\mathbf T_{H}\times_3\mathbf T_{S}\Big\|_F^2.
    \label{eq:fitting}
\end{equation}
We adopt ALS because the three mode unfoldings of \eqref{eq:parafac} are each linear in one factor,
\begin{align}
    \mathbf Y_{(1)}&=\mathbf T_A\mathbf Z_{A}^{T}+\mathbf N_{(1)},
    &\mathbf Z_{A}&=\mathbf T_{S}\kr\mathbf T_H,\label{eq:mode1}\\
    \mathbf Y_{(2)}&=\mathbf T_{H}\mathbf Z_{H}^{T}+\mathbf N_{(2)},
    &\mathbf Z_{H}&=\mathbf T_{S}\kr\mathbf T_{A},\label{eq:mode2}\\
    \mathbf Y_{(3)}&=\mathbf T_{S}\mathbf Z_{S}^{T}+\mathbf N_{(3)},
    &\mathbf Z_{S}&=\mathbf T_H\kr\mathbf T_{A},\label{eq:mode3}
\end{align}
where $\mathbf N_{(n)}$ is the corresponding unfolding of $\ten{N}$, so that each factor admits the closed-form conditional update
\begin{align}
    \widehat{\mathbf T}_{X}
    &=\arg\min_{\mathbf T_{X}}
      \big\|\mathbf Y_{(n)}-\mathbf T_{X}\widehat{\mathbf Z}_{X}^{T}\big\|_F^2
    \nonumber\\
    &=\mathbf Y_{(n)}\big(\widehat{\mathbf Z}_{X}^{T}\big)^{\dagger}
     =\mathbf Y_{(n)}\widehat{\mathbf Z}_{X}^{*}
      \big(\widehat{\mathbf Z}_{X}^{T}\widehat{\mathbf Z}_{X}^{*}\big)^{-1},
    \label{eq:ALS}
\end{align}
for $X\in\{A,H,S\}$ and $n=1,2,3$ respectively. Each update is exact given the other two factors, and the cycle requires no external optimization package. Two implementation choices matter. The cycle is initialized using the $L$ dominant left singular vectors of the corresponding unfolding, which provide a consistent estimate of each factor's column space in the noiseless case. Since \eqref{eq:fitting} is non-convex, the cycle is repeated from $N_{\rm s}-1$ further random initializations and the smallest-residual fit is retained; we use $N_{\rm s}=5$, tolerance $\epsilon=10^{-8}$ and $I_{\max}=200$.

\begin{remark}[Choice of solver]\label{rem:solver}
ALS is a block-coordinate method and converges only linearly. Gauss--Newton (GN) solvers for \eqref{eq:fitting} treat the three factors jointly and are the standard remedy when the model is ill-conditioned, in particular for the swamps and bottlenecks that arise when two components become nearly collinear in one or more modes. The TRACE framework\footnote{The code is available at \url{https://github.com/andrelimaferrer/TRACE-matlab}, archived at doi.org/10.5281/zenodo.22923602.} is agnostic to the choice: both solve \eqref{eq:fitting} and then proceed with the same recovery stage. ALS is adopted here because it is exact per block and, in the simulated scenarios of Section~\ref{sec:results}, is indistinguishable from the GN solver~\cite{tensorlab}.
\end{remark}

\subsection{Recovery of the physical parameters}\label{sec:postproc}
After the tensor-fitting stage, the estimated factors inherit the structure of \eqref{eq:THeq} and \eqref{eq:TAeq},
\begin{align}
\widehat{\mathbf{T}}_{H}&\approx\overline{\mathbf{F}}_{\mathrm{EM}}\overline{\mathbf{H}}_{\mathrm{EM}},
&
\widehat{\mathbf{T}}_{A}&\approx\mathbf{F}_{\mathrm{RF}}\mathbf{A}(\bm{\theta}),
\label{eq:TH_TA_relation}
\end{align}
so that each physical quantity is recovered by a sparse matching problem against the corresponding known training dictionary. Since every column of $\overline{\mathbf{H}}_{\mathrm{EM}}$ is a canonical basis vector, recovering it from $\widehat{\mathbf{T}}_{H}$ is a support-detection problem over the $M$ columns of $\overline{\mathbf{F}}_{\mathrm{EM}}$,
\begin{align}
\mathcal{P}_{\mathrm{EM}}\big(\widehat{\mathbf{T}}_{H},\overline{\mathbf{F}}_{\mathrm{EM}}\big)
&=\widehat{\overline{\mathbf{H}}}_{\mathrm{EM}}
=\big[\mathbf{e}_{\widehat{m}_{1}},\ldots,\mathbf{e}_{\widehat{m}_{L}}\big],
\label{eq:HEM_support_detection}\\
\widehat{m}_{r}&=\arg\max_{m}
\big|\overline{\mathbf{f}}_{\mathrm{EM},m}^{H}\widehat{\mathbf{t}}_{H,r}\big|,
\quad r=1,\ldots,L,\nonumber
\end{align}
where $\widehat{\mathbf{t}}_{H,r}$ is the $r$-th column of $\widehat{\mathbf{T}}_{H}$, $\overline{\mathbf{f}}_{\mathrm{EM},m}$ is the $m$-th column of $\overline{\mathbf{F}}_{\mathrm{EM}}$, and $\mathcal{P}_{\mathrm{EM}}$ the operator projecting the estimated EM factor onto the one-hot angular-index support set. The AoAs are recovered analogously, by a compressed beamspace search that matches each estimated steering vector n to the closest RF-projected one over an angular grid $\Theta$,
\begin{align}
\mathcal{P}_{A}\big(\widehat{\mathbf{T}}_{A},\mathbf{F}_{\mathrm{RF}}\big)
&=\widehat{\mathbf{A}}(\widehat{\bm{\theta}})
=\big[\mathbf{a}(\widehat{\theta}_{1}),\ldots,\mathbf{a}(\widehat{\theta}_{L})\big],
\label{eq:PA_operator}\\
\widehat{\theta}_{r}&=\arg\max_{\theta\in\Theta}
\big|\big(\mathbf{F}_{\mathrm{RF}}\mathbf{a}(\theta)\big)^{H}\widehat{\mathbf{t}}_{A,r}\big|,
\quad r=1,\ldots,L.\nonumber
\end{align}
\textcolor{black}{Because $\widehat m_r$ and $\widehat\theta_r$ are extracted from the same $r$-th PARAFAC component, they describe the EM index and AoA of the same path; the common permutation ambiguity only relabels the paths.}
Both problems are separable across $r$, so each costs a single pass over its dictionary, and no joint combinatorial search is required. This is the structural advantage of estimating the factors first: the greedy baseline of Section~\ref{sec:results} must search the \emph{joint} $G M$-atom dictionary, whereas \eqref{eq:HEM_support_detection} and \eqref{eq:PA_operator} decouple into one $M$-atom and one $G$-atom problem per component.

Projecting the detected supports back through the known training matrices gives the structured factors
\begin{align}
\widetilde{\mathbf{T}}_{H}&=\overline{\mathbf{F}}_{\mathrm{EM}}\widehat{\overline{\mathbf{H}}}_{\mathrm{EM}},
&
\widetilde{\mathbf{T}}_{A}&=\mathbf{F}_{\mathrm{RF}}\widehat{\mathbf{A}}(\widehat{\bm{\theta}}),
\label{eq:Ttilde}
\end{align}
which are then used to re-estimate the third factor by least squares,
\begin{equation}
\widehat{\mathbf{B}}=\widehat{\mathbf{T}}_{S}
=\mathbf{Y}_{(3)}
\big[(\widetilde{\mathbf{T}}_{H}\kr\widetilde{\mathbf{T}}_{A})^{T}\big]^{\dagger}.
\label{eq:Bhat_postprocessing}
\end{equation}
This final estimation steps re-references $\widehat{\mathbf{B}}$ to the true training matrices, thereby removing the scaling ambiguity. Finally, using $\mathbf B=\diag(\mathbf s)\mathbf C$, the stacked path-gain matrix follows from the structured LS problem
\begin{equation}
\widehat{\mathbf C}=\arg\min_{\mathbf C}
\big\|\widehat{\mathbf B}-\diag(\mathbf s)\mathbf C\big\|_F^2,
\label{eq:C_LS_problem}
\end{equation}
which, since $\diag(\mathbf s)$ is invertible whenever $s_p\neq0$, admits the exact closed-form solution $\widehat{\mathbf C}=\diag(\mathbf s)^{-1}\widehat{\mathbf B}$: each row of $\widehat{\mathbf B}$ is simply divided by the pilot symbol of the corresponding slot, with no combinatorial assignment involved. The complete procedure is summarized in Algorithm~\ref{alg:TRACE}.

\begin{algorithm}[!t]
\small
\caption{TRACE}
\label{alg:TRACE}
\setlength{\abovedisplayskip}{2pt}
\setlength{\belowdisplayskip}{2pt}
\renewcommand{\baselinestretch}{0.92}\selectfont
\begin{algorithmic}[1]
\REQUIRE $\ten{Y}\in\C^{N_{\mathrm{RF}}\times T\times P}$, $L$, $\overline{\mathbf{F}}_{\mathrm{EM}}$, $\mathbf{F}_{\mathrm{RF}}$, $\mathbf{s}$, grid $\Theta$.
\ENSURE  $\widehat{\mathbf{A}}(\widehat{\bm{\theta}})$, $\widehat{\overline{\mathbf{H}}}_{\mathrm{EM}}$, $\widehat{\mathbf{C}}$.
\STATE Initialize the factors from the $L$ dominant left singular vectors of the mode-$1$, mode-$2$ and mode-$3$ unfoldings.
\STATE Solve \eqref{eq:fitting} by cycling \eqref{eq:ALS} until the relative residual decrease falls below $\epsilon$ or $I_{\max}$ iterations elapse; repeat from $N_{\rm s}-1$ random initializations and keep the smallest residual.
\STATE Detect the EM support via \eqref{eq:HEM_support_detection} to obtain $\widehat{\overline{\mathbf{H}}}_{\mathrm{EM}}$.
\STATE Estimate the AoAs via \eqref{eq:PA_operator} to obtain $\widehat{\mathbf{A}}(\widehat{\bm{\theta}})$.
\STATE Form $\widetilde{\mathbf{T}}_{H}=\overline{\mathbf{F}}_{\mathrm{EM}}\widehat{\overline{\mathbf{H}}}_{\mathrm{EM}}$ and $\widetilde{\mathbf{T}}_{A}=\mathbf{F}_{\mathrm{RF}}\widehat{\mathbf{A}}(\widehat{\bm{\theta}})$.
\STATE Re-estimate $\widehat{\mathbf{B}}$ from \eqref{eq:Bhat_postprocessing}.
\STATE Recover $\widehat{\mathbf C}=\diag(\mathbf s)^{-1}\widehat{\mathbf B}$.
\RETURN $\widehat{\mathbf{A}}(\widehat{\bm{\theta}})$, $\widehat{\overline{\mathbf{H}}}_{\mathrm{EM}}$, $\widehat{\mathbf{C}}$.
\end{algorithmic}
\end{algorithm}

\section{Identifiability and Complexity}\label{sec:ident}
The model is identifiable up to permutation and scaling ambiguities of a PARAFAC decomposition. Both are immaterial, since each rank-one term corresponds to a single path, and reciprocal scalings leave the reconstructed tensor unchanged. Steps 3--6 of Algorithm~\ref{alg:TRACE} act column-wise and are permutation-equivariant, so no labeling is required. 
\begin{figure*}[!t]
\centering
\includegraphics[width=\textwidth]{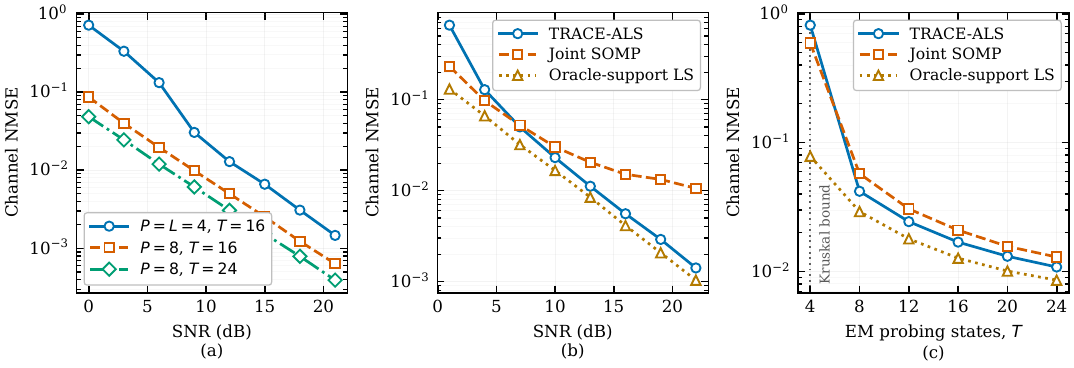}
\caption{Channel NMSE. (a) Training dimensions versus SNR, $N_{\rm r}=N_{\rm RF}=4$, $M=16$, $L=4$: pilot slots beyond the tensor rank remove the low-SNR threshold effect seen at $P=L$, while extra EM probing slots give a uniform gain. (b) Compressive configuration versus SNR, $N_{\rm r}=8$, $N_{\rm RF}=3$, $M=24$, $T=10$, $P=6$, $L=4$: the greedy dictionary baseline is more accurate below the crossover at about $7$~dB but saturates at a grid-induced floor, whereas TRACE keeps the SNR slope of the oracle-support bound. (c) EM probing states at $10$~dB, $N_{\rm r}=8$, $N_{\rm RF}=4$, $M=24$, $P=8$, $L=6$: satisfying Kruskal's condition with zero margin, at $T=4$, is not enough to make the model well conditioned.}
\label{fig:results}
\end{figure*}

Let $k_{H}$, $k_{A}$ and $k_{S}$ denote the Kruskal ranks of $\mathbf T_H$, $\mathbf T_{A}$ and $\mathbf T_{S}$. A sufficient uniqueness condition is Kruskal's~\cite{sidiropoulos2017tensor}
\vspace{-1ex}
\begin{equation}
\vspace{-1ex}
    k_H+k_{A}+k_{S}\geq 2L+2.
    \label{eq:kruskal}
\end{equation}
Assume that $\overline{\mathbf F}_{\rm EM}$ and $\mathbf F_{\rm RF}$ are drawn from well-conditioned codebooks, that the $L$ AoAs are distinct, the EM signatures create no repeated columns after projection, and the fading realizations are sufficiently diverse. Each factor then has Kruskal rank equal to its rank, $k_H=\min\{T,L\}$, $k_{A}=\min\{N_{\rm RF},L\}$ and $k_{S}=\min\{P,L\}$, and \eqref{eq:kruskal} becomes
\begin{equation}
    \min\{T,L\}+\min\{N_{\rm RF},L\}+\min\{P,L\}\geq 2L+2.
    \label{eq:system_parameter_condition}
\end{equation}
Uniqueness of the updates \eqref{eq:ALS} further requires the regression matrices $\mathbf Z_{A}\in\C^{PT\times L}$, $\mathbf Z_{H}\in\C^{PN_{\rm RF}\times L}$ and $\mathbf Z_{S}\in\C^{TN_{\rm RF}\times L}$ to have full column rank, which holds generically when $PT\geq L$, $PN_{\rm RF}\geq L$ and $TN_{\rm RF}\geq L$. These inequalities expose the trade-off among pilot slots, EM probing states and RF chains: the steering-factor update depends on $PT$, the EM-factor update on $PN_{\rm RF}$, and the fading update on $TN_{\rm RF}$.

\begin{remark}\label{rem:overdetermined}
Although \eqref{eq:fitting} may remain identifiable at $P=L$ and $T=L$, choosing overdetermined dimensions $P>L$ and $T>L$ markedly improves the conditioning and stability of the factor estimates, as Section~\ref{sec:results} confirms.
\end{remark}

The full-rank premise behind $k_H=\min\{T,L\}$ constrains the EM probing codebook and is easily violated by an apparently natural design. Subsampling an $M$-point DFT at the rows $\lfloor (t-1)M/T\rfloor+1$ makes columns $m$ and $m+T$ of $\overline{\mathbf F}_{\rm EM}$ \emph{identical} whenever $M/T$ is an integer, the omitted rows being exactly those that would distinguish them. 
Then $k_H$ collapses to $1$ and \eqref{eq:kruskal} fails. Randomized probing has no such structure, which is why it is used in Section~\ref{sec:results}. On the other hand, deterministic codebooks are safe when $T\geq M$.

\emph{Complexity.} Each ALS iteration is dominated by the three conditional updates in \eqref{eq:ALS} and scales as $\mathcal O(PTN_{\rm RF}L+L^3)$, so the cost of Step 2 is $N_{\rm s}$ times that of one restart. The recovery stage is of lower order: the EM support projection costs $\mathcal O(LMT)$ and the beamspace search $\mathcal O(LGN_{\rm RF})$ with $G=|\Theta|$ for a precomputed dictionary, while Step 7 is $\mathcal O(PL)$ divisions. 

\vspace{-2ex}
\section{Numerical Results}\label{sec:results}

The BS employs a uniform linear array of $N_{\rm r}$ pattern RAs, spaced at half-wavelength intervals, each cycling through $T$ reconfiguration states. For realization, the $L$ AoAs are drawn uniformly from $[-60^\circ, 60^\circ]$ subject to a minimum separation. Each path is associated with one EM angular-grid index and the gains are i.i.d.\ $\CN(0,1/L)$ across paths and slots. The pilot has constant modulus and unit energy. The entries of $\overline{\mathbf F}_{\rm EM}$ and $\mathbf F_{\rm RF}$ are unit modulus with independent uniform phases; both are drawn once and reused across realizations, reflecting a fixed training protocol. The SNR is the ratio of the average power of the noiseless pilot tensor to the noise power per entry. Performance is measured by the channel NMSE, $\sum_{p}\|\overline{\mathbf h}_{p}-\widehat{\overline{\mathbf h}}_{p}\|_2^2/\sum_{p}\|\overline{\mathbf h}_{p}\|_2^2$ with $\overline{\mathbf h}_{p}$ as in \eqref{eq:hbar}, which is invariant to a common permutation of the estimated columns.

Two baselines are used. A joint SOMP estimator~\cite{alkhateeb2014channel} operates on the mode-3 unfolding and greedily selects $L$ atoms from the joint dictionary $\{(\mathbf F_{\rm RF}\mathbf a(\theta))\kr(\overline{\mathbf F}_{\rm EM}\mathbf e_{m})\}$ indexed by $\Theta$ and the $T$ EM states, followed by an LS refinement on the selected support. An oracle-support LS reference is given the true $\mathbf A(\bm\theta)$ and $\overline{\mathbf H}_{\rm EM}$ and estimates only the gains; it is a lower bound, not a competing scheme. TRACE and SOMP share the same $1001$-point grid. All curves report the median over $200$ realizations: \eqref{eq:fitting} is non-convex, so at low SNR some realizations terminate at a stationary point with ${\rm NMSE}>1$, and a mean over such a mixture would measure the frequency of those failures. For Fig.~\ref{fig:results}(b) that fraction is $27\%$, $6\%$, and $0\%$ at $1$, $4$, and $7$~dB, and stays zero above, so outside the threshold region the two statistics coincide.

Fig.~\ref{fig:results}(a) examines performance in a non-compressive setting, $N_{\rm r}=N_{\rm RF}=4$, $M=16$, $T\geq M$, $L=4$ and $15^\circ$ minimum separation, so that the differences are attributable to the pilot and probing dimensions alone. All three configurations satisfy \eqref{eq:kruskal} with margin ($4+4+4\geq10$). We can see that raising $P$ improves the NMSE, as expected. In particular, settings with $P>L$ improve conditioning and robustness, whereas increasing $T$ from $16$ to $24$ yields a uniform improvement by about a factor of $1.6$.

Fig.~\ref{fig:results}(b) compares TRACE against the baselines in a compressive configuration, $N_{\rm r}=8$, $N_{\rm RF}=3$, $M=24$, $T=10$, $P=6$, $L=4$ and $5^\circ$ minimum separation. Both training matrices are now strictly wide, $N_{\rm RF}<N_{\rm r}$ and $T<M$, so the receiver observes a compressed projection in both domains, while \eqref{eq:kruskal} still holds ($4+3+4\geq10$). The two estimators exchange roles as the SNR grows. Below the crossover near $7$~dB, SOMP is more accurate by a factor of $2.8$ at $1$~dB: restricting each path to a dictionary atom is a hard structural constraint that suppresses noise where the unconstrained fit is poorly determined. Above the crossover, the same constraint becomes the binding limitation. SOMP improves by only $1.25$ per $3$~dB at high SNR, compared with the ideal factor of $2$ attained by both TRACE and the oracle, so its error flattens to a floor set by the dictionary's angular quantization and residual support-selection errors. In contrast, TRACE has no such floor, since it fits the factors in the continuum and consults the grid only once, in Step 4, on factors that are by then already accurate. For instance, at $20$~dB SNR, TRACE outperforms SOMP by almost one order of magnitude. The comparison against the oracle isolates what remains. At medium-to-high SNRs, TRACE tracks the bound at a nearly constant offset of $1.3$ to $1.4$, recovering the same SNR slope as an estimator given the true supports. Indeed, the residual offset is the price of estimating them and does not grow with SNR. 

Fig.~\ref{fig:results}(c) fixes the SNR at $10$~dB and sweeps the number of EM probing states, with $L=6$, $N_{\rm r}=8$, $N_{\rm RF}=4$, $M=24$, $P=8$ and $5^\circ$ minimum separation. The EM codebooks are nested, so a shorter design is a prefix of the longest one. 
Here \eqref{eq:system_parameter_condition} reads $\min\{T,6\}+4+6\geq14$ and is met from $T=4$ onward, the bound marked in the figure. The leftmost point therefore lies exactly at the sufficiency threshold, with zero margin. 
As in Fig.~\ref{fig:results}(b), the dictionary constraint makes SOMP the more accurate of the two in this ill-posed regime. Indeed, with four probing states, the $M=24$ columns of $\overline{\mathbf F}_{\rm EM}$ have mutual coherence $0.985$, so the support detection \eqref{eq:HEM_support_detection} cannot separate them, whereas at $T=24$ the coherence falls to $0.52$. Doubling to $T=8$ eliminates the failures outright and improves the NMSE. Beyond that, the returns diminish, while keeping a narrow gap to the oracle-support LS solution.

\section{Conclusion}
This letter proposes TRACE, a trilinear channel estimation framework for tri-hybrid architectures with reconfigurable antennas, that preserves the coupling among the EM angular response, the RF-projected steering matrix, and the pilot-weighted path gains. Representing the received pilots as a PARAFAC tensor yields a two-stage estimator that jointly fits the three factors and then maps them back to the physical channel parameters in closed form, with identifiability translated into explicit training design rules. Simulations confirm that TRACE retains the SNR slope of an oracle-support bound, while a dictionary-based greedy baseline, although more accurate at low SNR, saturates at a floor set by its grid. Extending the framework to the multi-user uplink is a natural perspective of this work.

\renewcommand{\baselinestretch}{.86}

\end{document}